\documentclass{article}
\usepackage{amsmath,amsfonts}
\usepackage{algorithmic}
\usepackage{algorithm}
\usepackage{array}
\usepackage[caption=false,font=normalsize,labelfont=sf,textfont=sf]{subfig}
\usepackage{textcomp}
\usepackage{stfloats}
\usepackage{url}
\usepackage{verbatim}
\usepackage{graphicx}
\usepackage{cite}
\usepackage{lineno,hyperref}
\modulolinenumbers[5]
\usepackage{amssymb}
\usepackage{latexsym}
\usepackage{wasysym}
\usepackage{amscd}
\usepackage{tikz-cd}

\newtheorem{thm}{Theorem}

\newtheorem{lem}{Lemma}
\newtheorem{prop}{Proposition}

\newtheorem{eg}{Example}

\newtheorem{defn}{Definition}

\newcommand {\pf}{\sc Proof.\ }
\newcommand {\epf}{$\dashv$}
\newcommand{\badstart}[0]{\ \\[-.2in]}
\newcommand{\norm}[1]{\|#1\|}
\title{Many-Valued Coalgebraic Modal Logic: One-step Completeness and Finite Model Property}
\author{Chun-Yu Lin and Churn-Jung Liau
\\ Institute of Information Science
\\ Academia Sinica, Nankang 115, Taipei, Taiwan
\\  email: liaucj@iis.sinica.edu.tw.}

\date{\, }

\begin{document}

\maketitle
\begin{abstract}
In this paper, we investigate the many-valued version of coalgebraic modal logic through predicate lifting approach. Coalgebras, understood as generic
transition systems, can serve as semantic structures for various kinds of modal logics. A well-known result in coalgebraic modal logic is that its completeness can be determined at the one-step level. We generalize the result to the finitely many-valued case by using the canonical model construction method. We prove the result for coalgebraic modal logics based on three different many-valued algebraic structures, including the finitely-valued {\L}ukasiewicz algebra, the commutative integral Full-Lambek algebra (FL$_{ew}$-algebra) expanded with canonical constants and Baaz Delta, and the FL$_{ew}$-algebra expanded with valuation operations. In addition, we also prove the finite model property of the many-valued coalgebraic modal logic by using the filtration technique.
\end{abstract}

\section{Introduction}
Coalgebraic modal logic, first proposed by Moss \cite{moss1999coalgebraic} in 1999, unifies various semantics of modal logics into a common framework using the theory of coalgebra \cite{jacobs2017introduction}. This  framework includes different class of models and many reasoning principles. Basically, there are two approaches toward coalgebraic modal logic--relation lifting \cite{moss1999coalgebraic} and predicate lifting\cite{pattinson2003coalgebraic}. The logic given by relation lifting approach, often called $\nabla$-logic,  encodes the modality in any set functor $T$ that preserves weak pullbacks. In this logic system, there are only one modal similarity type, namely the $\nabla$, in which the semantic is provided by the set functor $T$. However, the unusual syntax of this logic system makes it not easy to work with. For example, in \cite{venema2012completeness}, the authors proposed the first axiomatic system for $\nabla$-logic and use complicated techniques from coalgebra theory to prove its soundness and completeness. The second approach, predicate lifting, provides coalgebraic logic with a more standard modal syntax~\cite{pattinson2003coalgebraic}. But this requires a second parameter--the modal similarity type which is given by predicate lifting and not fixed in a logic system. Since its syntax is simpler than $\nabla$-logic, the proof of soundness and completeness of the logical system in \cite{pattinson2003coalgebraic} is less difficult and it can be addressed using one-step logic. In general, coalgebraic modal logic provides a uniform framework for a variety of applied modal logics developed in computer science and philosophy, including normal modal logic~\cite{blak,che}, minimal and monotone modal logics~\cite{che,pacuit}, graded modal logic~\cite{hoek92}, probability logic~\cite{halpern17,ognjanovic16}, and conditional logic~\cite{bur}.  A comprehensive survey of coalgebraic modal logic can be found in \cite{kupke2011coalgebraic}.

On the other hand, considering the modal reasoning in the context of vagueness and uncertainty, the fuzzy logic over residuated lattices (like that in \cite{hajek2013metamathematics}) appears as a suitable framework for developing logical systems. Thus, many-valued modal logic has been developed in response to the investigation of connection between modality and vagueness/uncertainty. In \cite{fitting1991many,fitting1992many}, Fitting presented a systematic study of many-valued modal logic using Heyting algebra and provided a multi-expert interpretation of the logic. Then, through abstract algebraic logic, Bou {\it et.al.} \cite{bou2011minimum} developed minimum many-valued modal logic over residuated lattices. While the many-valued modal logics considered in~\cite{bou2011minimum,fitting1991many,fitting1992many} are quite general, the formalization of the $[0,1]$-valued modal systems S5 and KD45 was also presented in~\cite{hajek2013metamathematics,hajek10}. These systems impose special constraints on the fuzzy accessibility relations of the Kripke models. For example, the fuzzy S5 system requires that the fuzzy accessibility relation is the universal relation. More generally, the framework for G\"odel modal logic has been also studied extensively. The $\Box$-fragment and $\Diamond$-fragment of different G\"odel modal systems, including K, D, T, S4, and S5, were axiomatized in~\cite{fml10}; and the full G\"odel modal logics K, T, S4, and S5 were axiomatized in~\cite{fml12}. Moreover, analytic proof methods for the $\Box$-fragment and $\Diamond$-fragment of the fuzzy K system, including sequent-of-relations and hypersequent calculi, were introduced in~\cite{lmcs11}. In addition to complete axiomatizations, the finite model property and the decidability of G\"odel modal logics were also investigated in~\cite{fml13,fml10,fml12,lmcs11}. Apart from the generalization of the Kripke semantics, many-valued modal logic based on neighborhood semantics was also proposed recently~\cite{Cintula18,Cintula19}. Moreover, the modal logic combining probability and fuzzy logic has been also explored in \cite{hajekGE95,Hajek07}.

It is then natural to expect that many-valued coalgebraic modal logic can unify different many-valued modal logics in the same way as in the two-valued case. This direction of research was studied by  B{\'\i}lkov{\'a}  and Dost{\'a}l both in relation lifting~\cite{bilkova2013many} and predicate lifting approaches~\cite{bilkova2016expressivity}. They showed that one can define many-valued semantics in both approaches and prove Hennessy-Milner property with some further assumptions. On the other hand, Schroeder and Pattinson use  coalgebraic techniques to study the complexity of the satisfiability problem in the context of fuzzy description logic, which can be regarded as a variant of fuzzy modal logic, under the infinitely-valued {\L}ukasiewicz semantics~\cite{SchroderP11}. However, these previous works do not touch upon the soundness and completeness of derivation systems for many-valued coalgebraic modal logic yet. Therefore, in this paper, we adopt the semantic given in \cite{bilkova2016expressivity} to prove that the soundness and completeness of many-valued coalgebraic modal logic can be determined at the one-step level. This generalizes a well-known result in coalgebraic modal logic to the finitely-valued case. In addition, we also show that the standard filtration method can be adapted to prove the finite model property of finitely-valued coalgebraic modal logic.

The rest of the paper is organized as follows. In Section~\ref{sec2}, we present preliminaries on many-valued logic and its algebraic semantics. In Section~\ref{sec3}, we present many-valued coalgebraic modal logic and its one-step fragments, including the syntax, semantics, and derivation system. In Section~\ref{sec4}, we prove the soundness and completeness theorem of full modal logic assuming the one-step soundness and completeness of the derivation system. In Section~\ref{sec5}, we use the filtration technique to establish the finite model property of the proposed logic. Finally, we summarize the result and indicate some directions for further work in Section~\ref{sec6}. Besides, we include appendices to review some basic notions of category used in this paper and exemplify some concrete instances of many-valued coalgebraic modal logic.

\section{Preliminaries on Many-Valued Logic}\label{sec2}
While truth values of the classical two-valued logic are always in the Boolean algebra ${\mathbf 2}=(\{0,1\},\wedge,\vee,\neg)$, many-valued logic is interpreted in different algebraic structures. In this paper, we consider three many-valued languages interpreted in lattice-based algebras. To introduce these algebraic structures, let us start with the commutative integral full-Lambek algebras (aka residuated lattice)\cite{Ono2003}.
\begin{defn}
	We say $\mathbb{A}= \langle A,\lor,\land,\to,\odot,0,1\rangle$ is a {\it commutative integral full-Lambek algebra (FL$_{ew}$-algebra)}  if
	\begin{itemize}
		\item $\langle A,\lor,\land,0,1 \rangle$ is a bounded lattice,
		\item $\langle A, \odot, 1 \rangle$ is a commutative monoid,
		\item Define the ordering $\leq$ over $A$ as $a\leq b$ iff $a \land b=b$ iff $a \lor b =a$,
		\item $\odot$ is residuated with $\to$, i.e. for all $a,b,c \in A$, $a\odot b \leq c$ iff $b \leq a \to c$,
		\item $0\leq a \leq 1$ for all $a \in A$.
	\end{itemize}
Here, $A$ is called the domain of the algebra $\mathbb{A}$. From now on, when we give an algebra $\mathbb{A}$, its domain is simply denoted by $A$.
\end{defn}

Because of the generality of the FL$_{ew}$-algebra, it covers a variety of truth value domains commonly used in the algebraic semantics of nonclassical logic. Some well-known special cases of FL$_{ew}$-algebras include Hetying algebras, MTL algebras, MV algebras, and BL algebras~\cite{bou2011minimum}. In particular, we are interested in MV chains with finite elements, or isomorphically, finitely-valued {\L}ukasiewicz algebras.
\begin{defn}
	An FL$_{ew}$-algebra $\mathbb{A}=\langle A,\lor,\land,\to,\odot,0,1\rangle$ is the $n$-valued MV chain (or {\L}ukasiewicz algebra) if
	\begin{itemize}
		\item $A=\{\frac{m}{n-1}\mid 0\leq m\leq n-1\}$,
        \item $a\wedge b=\min(a,b)$
        \item $a\vee b=\max(a,b)$
		\item $a\odot b=\max(0,a+b-1)$, and
		\item $a\to b=\min(1,1-a+b)$.
	\end{itemize}
We usually denote the $n$-valued {\L}ukasiewicz algebra by {\L}$_n$.
\end{defn}

We can extend the FL$_{ew}$-algebra with the Baaz Delta~\cite{baaz,hajek2013metamathematics} or valuation operations~\cite{maruyama11}. Let $\mathbb{A}= \langle A,\lor,\land,\to,\odot,0,1 \rangle$ be an FL$_{ew}$-algebra. Then, the Baaz Delta on $\mathbb{A}$ is an unary operation $\Delta:A\to\{0,1\}$ defined by \[
\Delta(x)=\left\{\begin{array}{ll}
1, & {\rm if} \; x=1;\\
0, & {\rm if} \; x\not=1.
\end{array}
\right.
\]
We call the algebra $\mathbb{A}_{\Delta}=(\mathbb{A},\Delta)$ a {\em $\Delta$-algebra\/}. In addition, the valuation operations $\tau_a:A\to\{0,1\}$ and $\upsilon_a:A\to\{0,1\}$ are defined for any $a\in A$ as follows:
\[
\tau_a(x)=\left\{\begin{array}{ll}
1, & {\rm if} \; x=a;\\
0, & {\rm if} \; x\not=a.
\end{array}
\right.
\]
\[
\upsilon_a(x)=\left\{\begin{array}{ll}
1, & {\rm if} \; x\geq a;\\
0, & {\rm if} \; x\not\geq a.
\end{array}
\right.
\]
The algebra $\mathbb{A}_{\tau\upsilon}=(\mathbb{A},(\upsilon_c)_{c\in A},(\tau_c)_{c\in A})$  is called a {\em $\tau\upsilon$-algebra\/} or a {\em valuation algebra\/}. Note that $\tau_1=\Delta$ and $\upsilon_a(x)=\bigvee_{c\geq a}\tau_c(x)$ for any $x\in A$ when $A$ is finite. We can also show that $\tau_a$ and $\upsilon_a$ are definable in {\L}$_n$ via the {\em strongly characterizing formula\/} $\eta_a$ introduced in \cite{bou2011minimum}. More specifically, the operation $\eta_a$ definable in {\L}$_n$ is the same as $\upsilon_a$ and $\tau_1=\eta_1$ and $\tau_a(x)=\eta_a(x)\wedge(\eta_{a'}(x)\to 0)$ if $a<1$, where $a'$ is the immediate successor of $a$ in the domain of {\L}$_n$.

Based on these algebraic notions, we can now introduce many-valued logic systems. Because we only consider finitely-valued logic in this paper, we assume the domain of our FL$_{ew}$-algebra is always finite hereafter. Let $\mathsf{P}$ be a set of propositional symbols and let $\mathbb{A}=\langle A,\lor,\land,\to,\odot,0,1\rangle$ be an FL$_{ew}$-algebra. The BNF syntax of our basic language ${\mathcal L}_\mathbb{A}$ is as follows\footnote{We identify a logical language with its set of (well-formed) formulas.}:
\[\varphi::=p\;|\;\bar{1}\;|\;\bar{0}\;|\;\varphi\lor\varphi\;|\;\varphi\land\varphi\;|\;\varphi\odot\varphi\;|\;\varphi\to\varphi,\] where $p\in \mathsf{P}$ is a propositional symbol. The connectives $\neg$ and $\leftrightarrow$ are defined as usual by $\neg\varphi:=\varphi\to\bar{0}$ and $\varphi\leftrightarrow\psi:=(\varphi\to\psi)\odot(\psi\to\varphi)$. We also usually write $\top$ and $\bot$ for $\bar{1}$ and $\bar{0}$ respectively. Note that we overload the logical connectives in the language with operations in the algebra for simplifying the presentation. This should not cause any confusion because we can always see the difference of their usages from the context. As an FL$_{ew}$-algebra can be extended with some unary operations, we can also enrich the basic language with corresponding connectives. Hence, we define the syntax of ${\mathcal L}_\Delta$ and ${\mathcal L}_{\tau\upsilon}$ by
\[\varphi::=p\;|\;\bar{c}\;|\;\Delta\varphi\;|\;\varphi\lor\varphi\;|\;\varphi\land\varphi\;|\;\varphi\odot\varphi\;|\;\varphi\to\varphi,\]
and
\[\varphi::=p\;|\;\bar{1}\;|\;\bar{0}\;|\;\tau_c(\varphi)\;|\;\upsilon_c(\varphi)\;|\;\varphi\lor\varphi\;|\;\varphi\land\varphi\;|\;\varphi\odot\varphi\;|\;\varphi\to\varphi,\]
respectively, where $c\in A$ is an element in the domain of the algebra. Note that, in the language ${\mathcal L}_\Delta$, we add a constant symbol, called the {\em canonical constant\/}, for each truth value besides the constants 0 and 1. This enhances the expressive power of the basic language as the valuation operations do. In fact, we can easily see that $\Delta(\bar{c}\to\varphi)$ and $\Delta(\bar{c}\leftrightarrow\varphi)$ exactly correspond to $\upsilon_c(\varphi)$ and $\tau_c(\varphi)$ respectively. However, ${\mathcal L}_\Delta$ is even more expressive than ${\mathcal L}_{\tau\upsilon}$ because the canonical constant $\bar{c}$ for $c\not=0,1$ is not definable in the latter. Nevertheless, including canonical constants in the language is not cost-free.  To preserve the completeness of the reasoning system while adding the extra expressive power, we need some book-keeping axioms, such as the axiom $\bigvee_{c\in A}(\varphi\leftrightarrow\bar{c})$ that exhaustively specifies possible truth values of any formula\cite{hajek2013metamathematics}. Then, to make the axiom finitary, we must restrict the domain of truth values to be finite. This is the main reason why we only consider finitely-valued logics in this paper.

In this paper, we will consider the languages $\mathcal{L}_{\text{\L}_n}$, $\mathcal{L}_\Delta$, and $\mathcal{L}_{\tau\upsilon}$ on respective algebras $\text{\L}_n$, $\mathbb{A}_\Delta$, and $\mathbb{A}_{\tau\upsilon}$. Hence, throughout this paper, we will use $\mathcal{L}$ on $\mathbb{A}$ to denote any one of the three languages when the definitions and results are applied to all of them so that it does not matter which language we specifically refer to. In addition, we sometimes use $\mathcal {L}(\mathsf{P})$ to denote the language when we need to emphasize that the language is constructed from the set of atomic formulas $\mathsf{P}$.

The formulas of these many-valued logic languages can be evaluated in the corresponding FL$_{ew}$-algebras. We define a {\em truth valuation\/} (or {\em truth assignment\/}) as a mapping $h:\mathsf{P}\to A$ and extend its domain to all formulas with homomorphism. In other words, $h(\bar{c})=c$ for any $c\in A$, $h(\ast\varphi)=\ast h(\varphi)$ for $\ast\in\{\Delta,\upsilon_c,\tau_c\mid c\in A\}$, and $h(\varphi\ast\psi)=h(\varphi)\ast h(\psi)$ for $\ast\in\{\wedge,\vee,\odot,\to\}$. Hence, we denote by $Hom({\mathcal L},\mathbb{A})$ the set of all truth assignments for a many-valued language ${\mathcal L}$ on an (extended) FL$_{ew}$-algebra $\mathbb{A}$. Let $\Gamma\cup\{\varphi\}$ be a set of formulas of the language ${\mathcal L}$ on $\mathbb{A}$. Then, $\varphi$ is said to be an $\mathbb{A}$-consequence of $\Gamma$, denoted by $\Gamma\models_\mathbb{A}\varphi$, if for all $h\in Hom({\mathcal L},\mathbb{A})$, $h[\Gamma]\subseteq\{1\}$ implies $h(\varphi)=1$. As in \cite{bou2011minimum}, we assume that we have a sound and complete axiomatization $\mathbf{Ax}(\mathbb{A})$ for the $\mathbb{A}$-consequence relation $\models_\mathbb{A}$ when we consider a many-valued language based on an algebra $\mathbb{A}$. While there exist finite residuated lattices that are not finitely axiomatizable, a lot of many-valued logics based on such algebras indeed have complete axiomatizations (for more details, see the appendix in \cite{bou2011minimum} and references therein.). Hence, the existence of complete axiomatizations for the underlying non-modal logics is a mild assumption here, even though it is somewhat restrictive.

\section{Many-Valued Coalgebraic Modal Logic}\label{sec3}
\subsection{Syntax}
As in the case of two-valued coalgebraic modal logic, the modalities in our language are also in 1-1 correspondence with {\em predicate liftings\/} for a set functor\footnote{See the appendix for the definition of functor and other basic notions of category theory.} $T$.
\begin{defn}
Let $T:\mathsf{Set} \to \mathsf{Set}$ be a set functor and let $\mathbb{A}$ be an (extended) FL$_{ew}$-algebra. Then, a predicate lifting is a natural transformation \[\lambda: Hom(-,A^n) \Rightarrow Hom(T(-),A),\] where $n\in\omega$ is called the arity of $\lambda$, recalling that $A$ is simply the domain of $\mathbb{A}$. When we specify the component of $\lambda$ with set $S$, its domain is a set of vector-valued functions from $S$ to $A^n$. Hence, we use the symbol $\langle f_1,\ldots,f_n \rangle$ where $f_i:S \to A$ for all $i=1,\ldots, n$ to denote such a vector-valued function.
\end{defn}
The modal language is then an extension of the many-valued language with modalities corresponding to predicate liftings. More specifically, let $\mathcal L$ be a many-valued language introduced in the preceding section and let $\Lambda$ be a set of  predicate liftings. Then the modal language $\Lambda(\mathcal{L})$ is the expansion of $\mathcal L$  with modal formulas defined by the following clause:
\begin{itemize}
\item if $\varphi_0,\ldots,\varphi_{n-1}$ are formulas in $\Lambda(\mathcal{L})$, then $\varhexagon_{\lambda}(\varphi_0,\ldots,\varphi_{n-1})$ is also a formula in $\Lambda(\mathcal{L})$ for any $n$-ary predicate lifting $\lambda\in\Lambda$.
\end{itemize}
The set of predicate liftings $\Lambda$ is called the {\em signature\/} of the modal language.

To study one-step logic, we need to define rank-0 and rank-1 fragments of the modal language. Let  $\Lambda$ be the signature of the modal language and let $T$ be its corresponding set functor. For any nonempty set $\Phi$, we define $T_{\Lambda}(\Phi)$ as the set
\[\{\varhexagon_{\lambda}(x_1,\ldots,x_n)\mid\lambda\in\Lambda, arity(\lambda)=n, x_1,\ldots,x_n\in\Phi\}.\]
Then, the {\em rank-0 and rank-1 fragments\/} of the modal language $\Lambda(\mathcal{L})$ are respectively $0\Lambda(\mathcal{L}):=\mathcal{L}$ and $1\Lambda(\mathcal{L}):=\mathcal{L}(T_{\Lambda}(\mathcal{L}))$, recalling our convention of using the notation $\mathcal{L}(\cdot)$ to explicitly indicate the set of atomic formulas for constructing the many-valued language. We typically use $\pi$ and $\alpha$ (possibly with subscripts) to denote rank-0 and rank-1 formulas respectively.

\subsection{Semantics}
The modal language is interpreted in coalgebraic structures. Recall that a $T$-coalgebra for a set functor $T$ is a pair $(S,\sigma)$ where $S$ is a nonempty set and $\sigma$ is a function from $S$ to $TS$.
\begin{defn}
Let $T$ be a set functor and let $\mathbb{A}$ be an (extended) FL$_{ew}$-algebra. Then a {\it T-model } on $\mathbb{A}$ is a triple $\mathbb{S}= \langle S,\sigma,V\rangle$, where $(S,\sigma)$ is a $T$-coalgebra and $V:\mathsf{P}\to Hom(S,A)$ is a truth valuation. We define the semantics $\norm{\cdot}_{\sigma}:S\to A$ of $\Lambda(\mathcal{L})$ formulas inductively: for all $s\in S$
\begin{itemize}
    \item $\norm{p}_{\sigma}:= V(p)$ for all $p\in \mathsf{P}$,   $\norm{\bar{c}}_{\sigma}(s):= c$ with $c\in A$,
    \item $\norm{\varphi\ast\psi}_{\sigma}(s):=\norm{\varphi}_{\sigma}(s)\ast\norm{\psi}_{\sigma}(s)$ for $\ast \in \{\lor,\land,\odot,\to\}$,
    \item $\norm{\ast\varphi}_{\sigma}(s):=\ast(\norm{\varphi}_{\sigma}(s))$ for $\ast \in \{\Delta,\tau_c,\upsilon_c\mid c\in A\}$,
    \item $\norm{\varhexagon_{\lambda}(\varphi_0,\ldots,\varphi_{n-1})}_{\sigma}(s):= \lambda_S(\langle \norm{\varphi_0}_{\sigma},\ldots,\norm{\varphi_{n-1}}_{\sigma} \rangle )(\sigma(s))$, where $\lambda\in\Lambda$ is an $n$-ary predicate lifting.
\end{itemize}
\end{defn}
In the current context, the carrier $S$ of a $T$-model denotes the set of states (aka.\ possible worlds). The notion of validity can then be defined as follows.

\begin{defn}
We say a formula $\varphi $ of $\Lambda(\mathcal{L})$ is {\it valid} in a $T$-model $\mathbb{S}$ if $\norm{\varphi}_{\sigma}(s)=1$ for all $s\in S$, and it is called {\it valid}, denoted by $\models_{\Lambda(\mathcal{L})}\varphi$, if it is valid in all $T$-models.
\end{defn}

While a $T$-model is a coalgebra with a truth valuation, we only need the unfolding of one single state to interpret rank-0 and rank-1 formulas. To do such an unfolding, we first introduce the notions of {\em marking\/} and {\em coloring\/}~\cite{schroder2010rank}. Let $\mathbb{A}$ be an (extended) FL$_{ew}$-algebra. Then we say a mapping $m:S\to Hom(\mathsf{P},A)$ is a {\it $\mathsf{P}$-marking} on $S$ and dually $m^{\flat}:\mathsf{P}\to Hom(S, A)$ is the coloring of $m$ if $m^{\flat}(p)(s) = m(s)(p)$ for any $s\in S$ and $p\in \mathsf{P}$. We can then define one-step frames and models based on markings and colorings.

\begin{defn}
Let $T$ be a set functor. A {\it one-step $T$-frame} is a pair $\langle S,\delta\rangle$ with $\delta \in TS$ . A {\it one-step $T$-model} over a set $\mathsf{P}$ of propositional symbols is a triple $\langle S,\delta, m\rangle$ such that $\langle S,\delta\rangle$ is a one-step $T$-frame and $m:S \to Hom(\mathsf{P}, A)$ is a $\mathsf{P}$-marking on $S$.
\end{defn}

\begin{defn}
Given a marking $m:S \to Hom(\mathsf{P},A)$, we define the 0-step interpretation $\norm{\pi}^0_m:S\to A$ of $\pi\in0\Lambda(\mathcal{L})$ recursively:  for all $s\in S$
\begin{itemize}
    \item $\norm{p}^0_m (s):= m^{\flat}(p)(s) $, $\norm{\bar{c}}^0_m(s) := c$ with $c \in A$,
    \item $\norm{\pi_0\ast\pi_1}^0_m (s):=\norm{\pi_0}^0_m(s)\ast\norm{\pi_1}^0_m(s)$ for $\ast\in\{ \lor,\land,\odot,\to\}$, and
    \item $\norm{\ast\pi}^0_m(s):=\ast\norm{\pi}^0_m(s)$ for $\ast\in\{\Delta,\tau_c,\upsilon_c|c\in A\}$.
\end{itemize}
\end{defn}
In addition, we can define the 1-step interpretation $\norm{\alpha}^1_m:TS\to A$ of $\alpha\in1\Lambda(\mathcal{L})$ in the following way.
\begin{defn}
Let $m$ be a marking and $\lambda$ is an $n$-ary predicate lifting. The 1-step interpretation of $\alpha\in1\Lambda(\mathcal{L})$ is defined as $$\norm{\varhexagon_{\lambda}(\pi_0,\ldots,\pi_{n-1})}^1_m(\delta) := \lambda_{S} (\langle \norm{\pi_0}^0_m,\ldots, \norm{\pi_{n-1}}^0_m \rangle )(\delta)$$ for any $\delta\in TS$ and standard clauses apply for many-valued logical connectives in the same manner as in $0\Lambda(\mathcal{L})$.
\end{defn}

\subsection{Derivation systems}
As usual, we can characterize the formal proof of validity in $\Lambda({\mathcal L})$ and its fragment with derivation systems.
\begin{defn}\badstart
\begin{enumerate}
\item A {\it logical rule} in the language $\Lambda(\mathcal{L})$ is a pair $\langle\Gamma,\varphi\rangle$, where $\Gamma\cup\{\varphi\}\subseteq\Lambda(\mathcal{L})$. If $\Gamma=\emptyset$, then $\varphi$ is called an {\em axiom}.
\item A {\it non-modal logical rule} $\langle\Gamma,\varphi\rangle$ is a logical rule where $\Gamma\cup\{\varphi\}\subseteq0\Lambda(\mathcal{L})$.
\item A {\it one-step logical rule} $\langle\Gamma,\varphi\rangle$ is a logical rule where $\Gamma\subseteq0\Lambda(\mathcal{L})$ and $\varphi\in1\Lambda(\mathcal{L})$.
\item A set of logical rules is called a {\em derivation system}. If a logical system contain only non-modal and one-step rules, it is called a {\em one-step derivation system}.
\end{enumerate}
\end{defn}
All derivation rules for the underlying many-valued logic are examples of non-modal logical rules. Examples of one-step logical rules are the congruence rule $C_{\lambda}$:
$$\frac{\pi_0\leftrightarrow \pi_0', \cdots,  \pi_{n-1} \leftrightarrow \pi_{n-1}'}{\varhexagon_{\lambda}(\pi_0,\ldots,\pi_{n-1}) \leftrightarrow \varhexagon_{\lambda}(\pi_0',\ldots,\pi_{n-1}')}$$
and the monotonicity rule $M_{\lambda}$
$$\frac{\pi_0\rightarrow \pi_0', \cdots,  \pi_{n-1} \rightarrow \pi_{n-1}'}{\varhexagon_{\lambda}(\pi_0,\ldots,\pi_{n-1}) \rightarrow \varhexagon_{\lambda}(\pi_0',\ldots,\pi_{n-1}')}$$ that we will associate with an $n$-ary modality $\varhexagon_{\lambda}$.

In general, logical rules represent a kind of schema so that we can instantiate them in the derivation process by using substitutions.
\begin{defn}
A {\it substitution} is a map $\rho:\mathsf{P}\to\Lambda(\mathcal{L})$. we will use the notation $(\varphi_i/p_i: p_i\in I)$ for the substitution that maps each variable $p_i\in I$ to the formula $\varphi_i$ and remain identical in other variables $p\in \mathsf{P}\backslash I$, where $I$ is a (typically finite) subset of $\mathsf{P}$. The application of a substitution $\rho=(\varphi_i/p_i: p_i\in I)$ to a formula $\varphi$ results in a new formula $\varphi\rho$ in which the variable $p_i$ is uniformly replaced by $\varphi_i$ for each $p_i\in I$.
\end{defn}

\begin{defn}
A {\it proof} or {\em derivation\/} of a formula $\varphi$ from a set of formula $\Gamma$ in a derivation system $\mathbf{L}$ is a well-founded tree (with no infinite branch) labeled by the formulas such that
\begin{itemize}
    \item its root is labeled by $\varphi$ and leaves by instances of axioms in $\mathbf{L}$ or elements of $\Gamma$ and
    \item if a node is labeled by $\psi$ and $\Phi\neq\emptyset$ is the set of labels of its preceding nodes, then $\langle\Phi,\psi\rangle$ is an instance of a rule in $\mathbf{L}$.
\end{itemize}
\end{defn}
We write $\Gamma\vdash_{\mathbf{L}}\varphi$ if there is a proof of $\varphi$ from $\Gamma$ in $\mathbf{L}$. If $\emptyset\vdash_{\mathbf{L}}\varphi$, then we say that $\varphi$ is {\it $\mathbf{L}$-derivable} and simply write it as $\vdash_{\mathbf{L}}\varphi$.

In this paper, our derivation systems $\mathbf{L}$ consist of (1) a set of logical rules  $R=\Gamma/\gamma$ where $\Gamma\subseteq\Lambda(\mathcal{L})$ and $\gamma\in\Lambda(\mathcal{L})$ (2) all axioms and rules from $\mathbf{Ax}(\mathbb{A})$ for the underlying many-valued logic, and (3) the congruence rule ($C_{\lambda}$) for each $\lambda\in\Lambda$.

\section{Soundness and Completeness}\label{sec4}
In this section, we are going to prove soundness and completeness theorem of many-valued coalgebraic modal logic from one-step logic by using the approach of canonical model construction~\cite{schroder2007finite}. First, we present the definition of one-step soundness and completeness.
\subsection{One-step logic}
\begin{defn}
A one-step logical rule $R=\langle\Gamma,\gamma\rangle$ where $\Gamma\subseteq0\Lambda(\mathcal{L})$ and $\gamma\in1\Lambda(\mathcal{L})$ is called {\it one-step sound} if for any marking $m$,  $\norm{\pi}^0_m = \norm{\top}^0_m$ for every $\pi\in\Gamma$ implies $\norm{\gamma}^1_m=\norm{\top}^1_m$. A one-step derivation system $\mathbf{L}$ is one-step sound if all of its one-step logical rules are one-step sound.
\end{defn}

\begin{defn}
We say $\pi\in0\Lambda(\mathcal{L})$ is a {\em true propositional fact\/} of a marking $m:S\to Hom(\mathsf{P},A)$ if $\norm{\pi}^0_m= \norm{\top}^0_m$. We use $TPF(m)$ to denote the set of true propositional facts of $m$, i.e.,
\[TPF(m):=\{\pi\in0\Lambda(\mathcal{L})\mid\norm{\pi}^0_m= \norm{\top}^0_m\}.\]
\end{defn}

\begin{defn}
A one-step derivation system $\mathbf{L}$ is {\it one-step complete} if for every marking $m:S\to Hom(\mathsf{P},A)$ and every $\alpha\in1\Lambda(\mathcal{L})$, we have $$\norm{\alpha}^1_m=\norm{\top}^1_m \mbox{ implies } TPF(m)\vdash_{\mathbf{L}}\alpha$$
\end{defn}

\subsection{Many-valued coalgebraic modal logic}
In this subsection, we will prove the main theorem of the paper. We start with the formal definition of soundness and completeness with respect to the full modal language.
\begin{defn}
Let $\mathbf{L}$ be a derivation system for the modal language $\Lambda(\mathcal{L})$. We say that $\mathbf{L}$ is {\it sound} if all $\mathbf{L}$-derivable formulas are valid, and {\it complete} if all valid formulas are $\mathbf{L}$-derivable. In notation, $\mathbf{L}$ is sound and complete if for any $\varphi\in\Lambda(\mathcal{L})$, $\models_{\Lambda(\mathcal{L})}\varphi$ iff $\vdash_{\mathbf{L}}\varphi$.
\end{defn}

\begin{defn}
Let $\Psi$ and $\Phi$ be two sets of $\Lambda(\mathcal{L})$-formulas. Then $\Psi$ is called the {\it closure} of $\Phi$ iff
\begin{enumerate}
    \item $\Phi \subseteq \Psi$
    \item $\Psi$ is closed under subformulas
    \item $\Psi$ contains $\top,\bot$ (and $\bar{c}$ for every $c\in A$ if $\mathcal{L}=\mathcal{L}_\Delta$).
\end{enumerate}
$\Phi$ is said to be {\it closed} if its closure is itself.
\end{defn}

As usual, it is easy to check the soundness. Hence, we only present the proof of completeness below. First, let us assume that $\Phi$ is a finite closed set of $\Lambda(\mathcal{L})$-formulas from now on. Then, we define
\[{\mathsf P}_\Phi:=\{a_{\varphi}:\varphi\in \Phi\}\] where $a_{\varphi}$ is a new propositional symbol for every $\varphi\in\Phi$, and use $0\Lambda(\mathcal{L}({\mathsf P}_\Phi))$ and $1\Lambda(\mathcal{L}({\mathsf P}_\Phi))$ to denote the rank-0 and rank-1 languages constructed from the set of new propositional symbols ${\mathsf P}_\Phi$.

In classical modal logic, the maximally consistent subsets of formulas play a crucial role in the canonical model construction. Analogously we employ non-modal homomorphisms to achieve the same purpose.
\begin{defn}
Let $\mathbb{A}$ be an (extended) FL$_{ew}$-algebra. Then, we say that a mapping $h:\Lambda({\mathcal L})\to A$ is a {\it non-modal homomorphism} if the following conditions hold:
\begin{itemize}
    \item $h(\bar{c})=c$ for any $c\in A$,
    \item $h(\varphi_1\ast\varphi_2)=h(\varphi_1)\ast h(\varphi_2)$ where $\ast\in\{\lor,\land,\odot,\to\}$,
    \item $h(\ast\varphi)=\ast h(\varphi)$ where $\ast\in\{\Delta,\tau_c,\upsilon_c\mid c\in A\}$.
\end{itemize}
\end{defn}
\noindent We often denote the set of non-modal homomorphisms as $Hom(\Lambda({\mathcal L}),\mathbb{A})$. Note that if $h\in Hom(\Lambda({\mathcal L}),\mathbb{A})$ is a non-modal homomorphism, then the restriction of $h$ to ${\mathcal L}$ is a truth valuation of the many-valued language ${\mathcal L}$, i.e., $h\!\!\upharpoonright\!\!{\mathcal L}\in Hom({\mathcal L},\mathbb{A})$. In fact, if  we construct a new many-valued language by extending $\mathsf{P}$ with the set of modal formulas, then $Hom(\Lambda({\mathcal L}),\mathbb{A})$ is simply the set of truth valuations of the extended language. More precisely, let $\mathsf{P}_\Lambda=\mathsf{P}\cup\{\varhexagon_{\lambda}(\varphi_1,\ldots,\varphi_n)\mid\lambda\in\Lambda, arity(\lambda)=n, \varphi_1,\ldots,\varphi_n\in\Lambda({\mathcal L})\}$. Then, $Hom(\Lambda({\mathcal L}),\mathbb{A})=Hom(\mathcal{L}(\mathsf{P}_\Lambda),\mathbb{A})$.

Let $\mathsf{Thm}$ denote the set of all  $\mathbf{L}$-derivable $\Lambda(\mathcal{L})$-formulas and let $\bar{S}$ denote the set of all non-modal homomorphisms that assign 1 to all formulas in $\mathsf{Thm}$. In notation,
\[\mathsf{Thm}=\{\varphi\in\Lambda(\mathcal{L})\mid\vdash_\mathbf{L}\varphi\},\]
\[\bar{S}=\{s\in Hom(\Lambda({\mathcal L}),\mathbb{A})\mid s[\mathsf{Thm}]\subseteq\{1\}\}.\]
Then, we have an analogy of the classical Lindenbaum Lemma.
\begin{lem}(Lindenbaum Lemma)
Let $\varphi\in\Lambda(\mathcal{L})$ be a formula in the modal language. Then, $\varphi\not\in \mathsf{Thm}$ implies that there exists $s\in\bar{S}$ such that $s(\varphi)<1$.
\end{lem}
{\pf}
Because the derivation system $\mathbf{L}$ contains $\mathbf{Ax}(\mathbb{A})$ for the underlying many-valued logic, we have that $\varphi$ is not $\mathbf{Ax}(\mathbb{A})$-derivable from $\mathsf{Thm}$ if it is not $\mathbf{L}$-derivable from $\mathsf{Thm}$ (i.e., $\varphi\not\in \mathsf{Thm}$). However, because $\mathbf{Ax}(\mathbb{A})$ characterizes the $\mathbb{A}$-consequence relation, this means that $\varphi$ is not an $\mathbb{A}$-consequence of $\mathsf{Thm}$. Hence, there exists $s\in Hom(\mathcal{L}(\mathsf{P}_\Lambda),\mathbb{A})=Hom(\Lambda({\mathcal L}),\mathbb{A})$ such that $s[\mathsf{Thm}]\subseteq\{1\}$ but $s(\varphi)<1$ by which the result follows immediately.
\epf

Next, we define the marking $m: \bar{S} \to Hom({\mathsf P}_\Phi,\mathbb{A})$ which maps s to $h_{s}\in Hom({\mathsf P}_\Phi,\mathbb{A})$ with $h_{s}(a_{\varphi})=s(\varphi)$. Then we have the following proposition.
\begin{prop}\label{prop1}
With the definition above, we can derive the following equality $$\norm{a_{\varphi}}^0_m (s)=s(\varphi) \mbox{ for any s} \in \bar{S}.$$
\end{prop}
{\pf}
This can be shown by $\norm{a_{\varphi}}_m^0 (s)=m^{\flat}(a_{\varphi})(s)=m(s)(a_{\varphi})=h_{s}(a_{\varphi})=s(\varphi)$.
\epf

Let $(\varphi/a_{\varphi}:\varphi\in\Phi)$ denote the natural substitution\footnote{We slightly abuse the notion of substitution here because it maps a formula from the language $\Lambda({\mathcal L}({\mathsf P}_\Phi))$ to one in another language $\Lambda({\mathcal L})$. By contrast, the standard definition of a substitution requires it to map a formula to another formula in the same language.} replacing all the variable $a_{\varphi}$ with the original formula $\varphi $. Therefore, for any formula $\pi \in 0\Lambda({\mathcal L}({\mathsf P}_\Phi))$ and $\alpha \in 1\Lambda({\mathcal L}({\mathsf P}_\Phi))$, we use $\hat{\pi}, \hat{\alpha} \in \Lambda({\mathcal L})$ respectively, to denote $\hat{\pi} := \pi(\varphi/a_{\varphi}: \varphi \in \Phi)$, and $\hat{\alpha} := \alpha (\varphi/a_{\varphi}: \varphi \in \Phi)$.

We then have the following lemma.

\begin{lem}(Stratification Lemma)
Let $\mathbf{L}$ be a one-step sound and complete derivation system for $1\Lambda({\mathcal L})$. Then,
\begin{enumerate}
\item For any formula $\pi\in 0\Lambda({\mathcal L}({\mathsf P}_\Phi))$, $\vdash_{\mathbf{L}}\hat{\pi}$ iff $\norm{\pi}^0_m(s)=1$ for any $s\in \bar{S}$;
\item For any formula $\alpha\in 1\Lambda({\mathcal L}({\mathsf P}_\Phi))$, $\vdash_{\mathbf{L}}\hat{\alpha}$ if $\norm{\alpha}^1_m(\delta)=1$ for any $\delta\in T\bar{S}$.
\end{enumerate}
\end{lem}
{\pf}
To prove the first part of this lemma, we first claim that $\norm{\pi}^0_m(s)=s(\hat{\pi})$ for any $\pi\in 0\Lambda({\mathcal L})({\mathsf P}_\Phi)$ and $s\in \bar{S}$.

This can be done by induction on the complexity of rank-0 formulas. For the base step, when $\pi$ is of the form  $a_{\varphi} \in {\mathsf P}_\Phi$, $\hat{\pi}$ is $\varphi$ by definition. Hence, $\norm{a_{\varphi}}_m^0(s)= s(\varphi)=s(\hat{\pi})$ using Proposition \ref{prop1}. The proof for $\bar{c}$ is straightforward by the definition of $\norm{\cdot}^0_m$ and $s$.

For the inductive step, first assume $\ast\in\{\lor,\land,\odot,\to\}$ and $\pi_1,\pi_2 \in 0\Lambda({\mathcal L})0({\mathsf P}_\Phi)$. By the inductive hypothesis, we have $$\norm{\pi_1\ast\pi_2}_m^0(s)=\norm{\pi_1}_m^0(s)\ast \norm{\pi_2}_m^0(s)=s(\hat{\pi_1})\ast s(\hat{\pi_2})=s(\hat{\pi_1}\ast \hat{\pi_2})=s(\widehat{\pi_1 \ast \pi_2}).$$ Second, let $\pi\in 0\Lambda({\mathcal L})({\mathsf P}_\Phi)$ and $\ast\in\{\Delta,\tau_c,\upsilon_c\mid c\in A\}$. $$\norm{\ast\pi}_m^0(s)=\ast\norm{\pi}_m^0(s)=\ast s(\hat{\pi})=s(\widehat{\ast\pi})$$ using the inductive hypothesis. This proves the claim.

Now, we prove the first part of the lemma. Suppose that $\vdash_{\mathbf{L}}\hat{\pi}$, i.e., $\hat{\pi}\in \mathsf{Thm}$. Hence, by the definition of $\bar{S}$, we have $s(\hat{\pi})=1$ for all $s\in \bar{S}$. Using the above claim,  $$\norm{\pi}^0_m(s)=s(\hat{\pi})=1$$ which proves the result we want. On the other hand, assume that $\nvdash \hat{\pi}$. Then $\hat{\pi}\notin \mathsf{Thm}$. Thus, there exists an $s\in\bar{S}$ such that $s(\hat{\pi})\not=1$ by the Lindenbaum Lemma. By the claim again, we have $\norm{\pi}^0_m(s)=s(\hat{\pi}) \neq 1$.

For the proof of the second part of the lemma, we first note that, because we treat axioms and rules as schemata, a derivation system and its (one-step) soundness and completeness are independent of its choice of the underlying propositional variables. Hence, $\mathbf{L}$, as a one-step sound and complete derivation system for $1\Lambda({\mathcal L})$, is also one-step complete for $1\Lambda({\mathcal L}({\mathsf P}_\Phi))$. Hence, suppose that $\norm{\alpha}^1_m(\delta)=1$ for any $\delta\in T\bar{S}$. We have $TPF(m)\vdash_{\mathbf{L}}\alpha$ by the definition of one-step completeness, i.e. there is a derivation of $\alpha$ from $TPF(m)$. We prove $\vdash_\mathbf{L}\hat{\alpha}$ by induction over the complexity of the derivation.

Tracing the tree structure of the proof, the leaves are either axioms of $\mathbf{L}$ or axioms of ${\mathbf Ax}(\mathbb{A})$ or elements of $TPF(m)$. For the base case, we only need to consider $\pi\in TPF(m)$. However, the first part of the lemma has implied that $\vdash_{\mathbf{L}}\hat{\pi}$. Then, attaching the $\mathbf{L}$-proof of $\hat{\pi}$ to the derivation tree of $TPF(m)\vdash_{\mathbf{L}}\alpha$ and substituting $a_\varphi$ with $\varphi$ uniformly lead to an $\mathbf{L}$-proof of $\hat{\alpha}$.
\epf

\begin{defn}
Let $\bar{S}$ be defined as above. We define a syntactical evaluation  $|\varphi|:\bar{S} \to \mathbb{A}$ for every $\varphi\in \Lambda({\mathcal L})$ as $s \mapsto s(\varphi)$.
\end{defn}
This is clearly a well-defined function. Thus, for any $\varphi\in\Phi$, we have $$\norm{a_{\varphi}}^0_m = |\varphi|$$ since $\norm{a_{\varphi}}^0_m(s)=s(\varphi)=|\varphi|(s)$ for every $s \in \bar{S}$ by definition. Now, we prove the key lemma that will be used in the proof of completeness theorem.

\begin{lem}(Existence Lemma)
Let $\Phi$ be given as above. Then, there exists a map $\sigma: \bar{S} \to T\bar{S}$ such that for all $s \in \bar{S}$ and all formulas of the following form  $\varhexagon_{\lambda}(\varphi_0,\ldots,\varphi_{n-1}) \in \Phi$, we have $$s(\varhexagon_{\lambda}( \varphi_0,\ldots,\varphi_{n-1}))= \lambda_{\bar{S}}(\langle |\varphi_0|,\ldots,|\varphi_{n-1}| \rangle )(\sigma(s))$$
\end{lem}
{\pf} We prove the lemma by considering $\mathcal{L}=\mathcal{L}_{\text{\L}_n}, \mathcal{L}_\Delta$, and ${\mathcal L}_{\tau\upsilon}$ respectively.

First, in the case of $\mathcal{L}_\Delta$, assume for a contradiction that for some $s\in \bar{S}$ there is no $\sigma(s)$ satisfying the equality. Let us fix such an $s$ in the proof below.  List all formulas of the form $\varhexagon_{\lambda_i}(\varphi_0,\ldots,\varphi_{n_i-1})$ and denote them as $\varhexagon_{\lambda_0}\psi_0,\ldots, \varhexagon_{\lambda_{k-1}}\psi_{k-1}$ if the number of such formulas is $k$. Each $\psi_i$ is in fact a list $(\varphi_0,\ldots,\varphi_{n_i-1})$ of length $n_i$ if $\lambda_i$ is an $n_i$-ary predicate lifting. Suppose that $s(\varhexagon_{\lambda_i}(\psi_i))=c_i\in A$ for all $i=0,\ldots,k-1$. Then, the assumption for $s$ implies that for any $\delta\in T\bar{S}$, there exists $i$ such that $(\lambda_i)_{\bar{S}}(|\psi_i|)(\delta)\neq c_i$.

Next, we define the following formula $\alpha\in 1\Lambda({\mathcal L}({\mathsf P}_\Phi))$,
\[\alpha := \Delta(\bigwedge_{i=0}^{k-1}(\varhexagon_{\lambda_i}a_{\psi_i}\leftrightarrow \bar{c}_i))\to\bot.\]
Note that each propositional symbol $a_{{\psi}_i}$ is actually a list $(a_{\varphi_0},\ldots,a_{\varphi_{n_i-1}})$ of length $n_i$ whenever $\lambda_i$ is an $n_i$-ary predicate lifting. We write it as $a_{\psi_i}$ to simplify the notation. Then for any $\delta\in T\bar{S}$
\begin{eqnarray*}
    \norm{\alpha}^1_m(\delta) & =& \norm{\Delta(\bigwedge_{i=0}^{k-1}(\varhexagon_{\lambda_i}a_{\psi_i}\leftrightarrow \bar{c}_i)) \to \bot}^1_m(\delta)  \\
    &=& \norm{\Delta(\bigwedge_{i=0}^{k-1}(\varhexagon_{\lambda_i}a_{\psi_i}\leftrightarrow \bar{c}_i))}^1_m(\delta) \to \norm{\bot}^1_m(\delta)\\
    & =& \Delta(\norm{(\bigwedge_{i=0}^{k-1}(\varhexagon_{\lambda_i}a_{\psi_i}\leftrightarrow \bar{c}_i))}^1_m(\delta)) \to \norm{\bot}^1_m(\delta)\\
    &=& \Delta(\bigwedge_{i=0}^{k-1}((\lambda_i)_{\bar{S}}(\norm{a_{\psi_i}}^0_m)(\delta))\leftrightarrow c_i) \to 0 \\
    &= & \Delta(\bigwedge_{i=0}^{k-1}((\lambda_i)_{\bar{S}}(|\psi_i|)(\delta))\leftrightarrow c_i) \to 0
\end{eqnarray*}
Since for any $\delta\in T\bar{S}$, there exists $i$ such that $(\lambda_i)_{\bar{S}}(|\psi_i|)(\delta)\neq c_i$ by our assumption, we have  $$\bigwedge_{i=0}^{k-1}((\lambda_i)_{\bar{S}}(|\psi_i|)(\delta))\leftrightarrow c_i) \neq 1,$$ for any $\delta\in T\bar{S}$. That is, $\norm{\alpha}^1_m(\delta)= 0 \to 0 =1$ for any $\delta\in T\bar{S}$. From the Stratification Lemma, we get $\vdash_\mathbf{L}\hat{\alpha}$ which means $s(\hat{\alpha})=1$. However, compute $s(\hat{\alpha})$ gives us
\begin{eqnarray*}
    s(\hat{\alpha}) & =& s(\Delta(\bigwedge_{i=0}^{k-1}(\varhexagon_{\lambda_i}\psi_i\leftrightarrow \bar{c}_i)) \to \bot)\\
    &=& \Delta(\bigwedge_{i=0}^{k-1}s(\varhexagon_{\lambda_i}\psi_i\leftrightarrow \bar{c}_i)) \to 0\\
    &=& \Delta(1) \to 0 \\
    &=&0
\end{eqnarray*}
This contradicts to the value of $\hat{\alpha}$ we have above.

Second, for the case of ${\mathcal L}_{\tau\upsilon}$, the proof is similar except that $\alpha$ is defined as
\[\alpha:=(\bigwedge_{i=0}^{k-1}\tau_{c_i}(\varhexagon_{\lambda_i}a_{\psi_i}))\to\bot.\]

Finally, in the case of $\mathcal{L}_{\mbox{\L}_n}$, because $\tau_c$ is definable in $\mbox{\L}_n$ for any $c\in \mbox{\L}_n$, we can show that there exists a formula $\alpha$ in $1\Lambda(\mathcal{L}_{\mbox{\L}_n}({\mathsf P}_\Phi))$ that leads to the contradiction as in the case of ${\mathcal L}_{\tau\upsilon}$.
\epf

\begin{lem}(Truth Lemma)
Let $\Phi$ be the closed set of $\Lambda({\mathcal L})$-formulas defined above and let $\sigma:\bar{S} \to T\bar{S}$ be the map satisfying the Existence Lemma. Then we have $$ s(\varphi)=\norm{\varphi}_{\sigma}(s)$$ for all $s\in \bar{S}$ and $\varphi\in \Phi$
\end{lem}
{\pf}
Let $\mathbb{S}$ be a $T$-model given by $\sigma$ as above and a valuation $V:\mathsf{P}\to Hom(\bar{S},A)$ defined as $V(p)(s)=s(p)$ for $s\in\bar{S}$ and $p\in \mathsf{P}$. We prove this lemma by a straightforward induction on the complexity of formulas.
\begin{enumerate}
\item the base case: $s(p)=V(p)(s)=\norm{p}_{\sigma}(s)$ for any $p\in \mathsf{P}$ and $s(\bar{c})=c=\norm{\bar{c}}_{\sigma}$ for any $c\in A$.
\item for the logical connectives:
\[s(\varphi\ast\psi)=s(\varphi)\ast s(\psi)=\norm{\varphi}_{\sigma}(s)\ast\norm{\psi}_{\sigma}(s) = \norm{\varphi\ast\psi}_{\sigma}(s)\] for $\ast \in\{\lor,\land,\odot,\to\}$.
\item for the unary operations: $s(\ast\varphi)=\ast(s(\varphi))=\ast(\norm{\varphi}_{\sigma}(s))= \norm{\ast\varphi}(s)$ for $\ast \in\{\Delta,\tau_c,\upsilon_c\mid c\in A\}$.
\item for any $n$-ary predicate lifting $\lambda\in\Lambda$, \begin{eqnarray*}
    s(\varhexagon_{\lambda}(\varphi_0,\ldots,\varphi_{n-1})) & =&\lambda_{\bar{S}}(\langle |\varphi_0|,\ldots,|\varphi_{n-1}| \rangle)(\sigma(s)) \\
    & =& \lambda_{\bar{S}}(\langle \norm{\varphi_0}_{\sigma},\ldots,\norm{\varphi_{n-1}}_{\sigma} \rangle )(\sigma(s))\\
    &=& \norm{\varhexagon_{\lambda}(\varphi_0,\ldots,\varphi_{n-1})}_{\sigma}(s)
\end{eqnarray*}
\end{enumerate}
Note that the first equality is given by the Existence Lemma and this is the main reason why the result only holds for formulas in $\Phi$. Also, the second equality is by the inductive hypothesis $s(\varphi)=\norm{\varphi}_{\sigma}(s)$ and the definition $|\varphi|(s)=s(\varphi)$.
\epf

Now we are ready for proving the soundness and completeness theorem.
\begin{thm}
Let $\Lambda$ be a modal signature for the set functor $T$ and let $\mathbf{L}$ be a one-step sound and complete system for $1\Lambda({\mathcal L})$. Then $\mathbf{L}$ is also sound and complete with respect to the full modal language $\Lambda({\mathcal L})$.
\end{thm}
{\pf}
To prove the soundness, it is easy to check that all axioms of $\mathbf{L}$ are valid and all its logical rules preserve the validity. For the proof of completeness, suppose that $\psi$ is a $\Lambda({\mathcal L})$-formula not derivable from the system $\mathbf{L}$, i.e., $\psi\not\in \mathsf{Thm}$. Take $\Psi$ to be the closure of $\{\psi\}$. Then, by the Lindenbaum Lemma, there exists an $s\in\bar{S}$ such that $s(\psi)<1$. Therefore, by using the $T$-model $\langle\bar{S},\sigma,V\rangle$ defined in the proof of truth lemma, we have $\norm{\psi}_{\sigma}(s)=s(\psi)<1$ which implies that $\psi$ is not valid.
\epf

\section{Filtration and Finite Model Property}\label{sec5}
In the preceding section, we prove the completeness theorem by using the canonical model construction. However, because the canonical model may have infinite number of states, the finite model property does not follow as a corollary of the construction as in the classical case\cite{schroder2007finite}. Hence, in this section, we employ the filtration method~\cite{blak,che} to establish the finite model property of the above-mentioned logics. To start, we notice that, although the canonical model constructed above may contain an infinite number of states, the truth lemma only holds for a finite closed set of formulas that we are interested in. In other words, it does not matter whether two states different in the truth values of formulas outside that set. Filtration method is a standard technique to make the idea explicit by identifying all states that coincide on the truth values of a finite set of formulas.

Formally, we can define the filtrated model for our logics by slightly modifying that for classical coalgebraic modal logic\footnote{The filtration method for classical coalgebraic modal logic was presented in an unpublished tutorial introduction by Y. Venema.}. First, let $\Phi$ be a finite closed set of $\Lambda(\mathcal{L})$-formulas as defined above and let $\mathbb{S}= \langle S,\sigma,V\rangle$ be a $T$-model. Then, we can define an equivalence relation $\equiv_\Phi\subseteq S\times S$  by
\[s\equiv_\Phi t\;\;\mbox{iff for all}\;\; \varphi\in\Phi, \norm{\varphi}_\sigma(s)=\norm{\varphi}_\sigma(t).\] For convenience, we denote the $\equiv_\Phi$-equivalence class containing a state $s$ as $\underline{s}$ and define the quotient set and quotient map as $\underline{S}=\{\underline{s}\mid s\in S\}$ and $q:S\to\underline{S}$ such that $q(s)=\underline{s}$ respectively. Next, we define a representative choice function $r:\underline{S}\to S$ as a pseudo-inverse of $q$ which selects an arbitrary element $r(\underline{s})$ from each equivalence class $\underline{s}$. Then, the coalgebra map $\underline{\sigma}:\underline{S}\to T\underline{S}$ is defined by $\underline{\sigma}:= Tq\circ\sigma\circ r$. That is, $\underline{\sigma}$ is chosen to make the diagram below commutative.
\begin{center}
\begin{tikzcd}
\underline{S} \arrow[r,shift left,"\underline{\sigma}"] \arrow[d,"r"]& T\underline{S}\\
  S \arrow[u,shift right,"q"] \arrow[r,"\sigma"] & TS \arrow[u,"Tq"]\\
\end{tikzcd}
\end{center}

In summary, we have the formal definition of the $\Phi$-filtration of $T$-models as follows.
\begin{defn}
Let $\Phi$ be a finite closed set of $\Lambda(\mathcal{L})$-formulas and let $\mathbb{S}= \langle S,\sigma,V\rangle$ be a $T$-model. Then, a $\Phi$-filtration of $\mathbb{S}$ is any $T$-model $\underline{\mathbb{S}}= \langle\underline{S},\underline{\sigma},\underline{V}\rangle$ such that
\begin{enumerate}
\item $\underline{S}$ is the quotient set of $S$ with respect to the equivalence relation $\equiv_\Phi$,
\item $\underline{\sigma}:= Tq\circ\sigma\circ r$ for some representative choice function $r$, and
\item the truth valuation $\underline{V}$ must satisfy the condition that $\underline{V}(p)(\underline{s})=V(p)(s)$ for any atomic formula $p\in\Phi$ and $\underline{s}\in\underline{S}$.
\end{enumerate}
\end{defn}
Note that a $T$-model may have more than one filtrations. In particular, different representative choice functions may result in different filtrations for the same $T$-model. However, all $\Phi$-filtration of a given $T$-model enjoy a common property, that is, they all preserve the truth values of formulas in $\Phi$. This is formally stated as the following filtration lemma.
\begin{lem}(Filtration Lemma)
Let $\Phi$ be a finite closed set of $\Lambda({\mathcal L})$-formulas and let $\underline{\mathbb{S}}= \langle\underline{S},\underline{\sigma},\underline{V}\rangle$ be a $\Phi$-filtration of the $T$-model $\mathbb{S}= \langle S,\sigma,V\rangle$. Then
\[\norm{\varphi}_\sigma(s)=\norm{\varphi}_{\underline{\sigma}}(\underline{s})\]
for any $\varphi\in\Phi$ and $s\in S$.
\end{lem}
{\pf} The proof is a simple induction on the structure of formulas. The inductive base holds by the requirement imposed on the truth valuation $\underline{V}$ and the inductive steps for non-modal cases are fairly straightforward. Hence, let us proceed with the case of modal formulas, i.e., $\varphi=\varhexagon_{\lambda}\psi$, where, for simplification, we assume that $\lambda$ is an unary predicate lifting and the proof is easily generalized to modalities of arbitrary arities.

Recall that $\lambda$ is a natural transformation between two contravariant functors $Hom(-,A)$ and $Hom(T(-),A)$. The functor $Hom(-,A)$ (and similarly $Hom(T(-),A)$) sends a set $X$ to the set $Hom(X,A)$ of all functions from $X$ to $A$, called a Hom-set, and send a morphism $f:X\to Y$ to a morphism $Hom(f,A):Hom(Y,A)\to Hom(X,A)$ defined by $Hom(f,A)(g)=g\circ f$ for any $g\in Hom(Y,A)$. Note that a contravariant functor reverses the direction of a morphism $X\to Y$ to $Hom(Y,A)\to Hom(X,A)$. In the following paragraphs, to simplify the notation, we will omit the underlying domain of truth values and simply write these two functors as $H(-)$ and $HT(-)$ respectively. Hence, according to the semantics of the logic, what we have to prove is
\[\norm{\varphi}_\sigma=\norm{\varphi}_{\underline{\sigma}}\circ q=H(q)\norm{\varphi}_{\underline{\sigma}}\]
or equivalently,
\[\norm{\varphi}_{\underline{\sigma}}=\norm{\varphi}_\sigma\circ r=H(r)\norm{\varphi}_\sigma\]
for any $\varphi\in\Phi$. In both equations, the left equality is simply a restatement of what we want to prove and the right one arises from the definition of the functor $H$.

Then, the inductive step proceeds as follows.
\[\begin{array}{lcll}
\norm{\varhexagon_{\lambda}\psi}_{\underline{\sigma}}&=&\lambda_{\underline{S}}(\norm{\psi}_{\underline{\sigma}})\circ\underline{\sigma}& \mbox{semantics of}\; \varhexagon_{\lambda}\\
&=&H(\underline{\sigma})\lambda_{\underline{S}}(\norm{\psi}_{\underline{\sigma}})&\mbox{definition of}\; H(\underline{\sigma})\\
&=&H(r)H(\sigma)HT(q)\lambda_{\underline{S}}(\norm{\psi}_{\underline{\sigma}}) & \mbox{definition of}\; \underline{\sigma}\\
&=&H(r)H(\sigma)\lambda_SH(q)(\norm{\psi}_{\underline{\sigma}}) & \mbox{naturality of}\; \lambda\\
&=&H(r)H(\sigma)\lambda_S(\norm{\psi}_\sigma) & \mbox{induction hypothesis}\\
&=&H(r)(\lambda_S(\norm{\psi}_\sigma)\circ\sigma) & \mbox{definition of}\; H(\sigma)\\
&=&H(r)\norm{\varhexagon_{\lambda}\psi}_\sigma & \mbox{semantics of}\; \varhexagon_{\lambda}\\
\end{array}
\]
We can summarize the inductive step in the following diagram.
\begin{center}
 \begin{tikzcd}
H\underline{S} \arrow[d,"Hq"]\arrow[r,"\lambda_{\underline{S}}"]& HT\underline{S} \arrow[d,"HTq"] \arrow[r,"H\underline{\sigma}"]& H\underline{S}\\
  HS  \arrow[r,"\lambda_S"] & HTS \arrow[r,"H\sigma"]& HS \arrow[u, "Hr"]\\
\end{tikzcd}   
\end{center}
\epf

From the filtration lemma, we can easily derive the finite model property of our logic.
\begin{thm}(Finite Model Property)
Let $\varphi$ be a $\Lambda({\mathcal L})$-formula, where ${\mathcal L}$ is a $k$-valued logic. If $\varphi$ is satisfiable in a $T$-model, then it is satisfiable in a finite $T$-model $\mathbb{S}= \langle S,\sigma,V\rangle$ such that $|S|$ is $O(k^{|\varphi|})$, where $|\varphi|$ denotes the size of the formula.
\end{thm}
{\pf}Let $\Phi$ be the smallest closed set containing $\varphi$. Then, the cardinality of $\Phi$ is at most $|\varphi|+k$. Hence, by the filtration lemma, we have the $\Phi$-filtration of the model satisfying $\varphi$ with size at most $k^{|\Phi|}$, which is obviously $O(k^{|\varphi|})$. \epf

\section{Concluding Remarks}\label{sec6}
In this paper, we prove that there is a sound and complete derivation system for finitely-valued coalgebraic modal logic under the assumption of one-step soundness and completeness. Hence, we generalize a well-known result of coalgebraic modal logic to the many-valued case. That is, the soundness and completeness of coalgebraic modal logic can be determined at the one-step level. In addition, we also adapt the classical filtration method to prove the finite model property of the proposed logic. The main contribution of the paper is the proof of such results for finitely-valued coalgebraic modal logic. Admittedly, the structure of the proof based on the canonical model construction is similar to that in the two-valued case. However, there are remarkable differences on the technical details of the proof. In particular, a crucial difference is on the proof of the Existence Lemma. In the proof, we need to design a witness formula $\alpha$ to derive the existence of the coalgebraic map on the set of states. For the classical case, a state $s$ in the canonical model is simply a maximally consistent subset of formulas. Thus, $\alpha$ is easily defined as the conjunction of rank-1 modal formulas in $s$ and the negations of rank-1 modal formulas not in $s$. However, in the many-valued case, a state $s$  in the canonical model is a non-modal homomorphism from the set of formulas to the underlying algebra of truth values. Hence, the witness formula needs to be able to ``read out'' the truth value of every rank-1 modal formula in $s$. We achieve this purpose by using the combination of canonical constants and Baaz Delta or valuation operations. In this way, we provide a novel method for the canonical model construction of finitely-valued coalgebraic modal logic.

We consider modal logics based on three different many-valued language. A common feature of these languages is that we can internalize the meta-level valuation operations into the language. Consequently, we can express the truth value of any formula in the language which plays a crucial role in the proof of the key lemma for the canonical model construction as mentioned above. However, not all many-valued languages have the property. Hence, how to extend our result to many-valued coalgebraic modal logic without such property is an important issue to be addressed in the next step. In fact, in addition to the approach adopted here, an alternative method based on inductive construction  is also popular in proving the completeness of coalgebraic modal logic~\cite{pattinson2003coalgebraic}. Therefore, it is likely to use the inductive method to prove the completeness of many-valued coalgebraic modal logic that cannot express the truth values of its formulas in the object language.

In addition, another pressing issue is to provide a concrete instance of one-step sound and complete logic system. In the classical case, this can be done using the equivalent concept of one-step consistency and satisfiable \cite{schroder2010rank}. However, there is no corresponding definition of such equivalence in many-valued logic. Thus, showing one-step soundness and completeness of a derivation system becomes more difficult in the many-valued case. One possible exemplary one-step complete system is the fuzzy two-layered modal logic proposed in \cite{baldi2020classical,cintula2014modal}. We can restrict the syntax and language of in \cite{cintula2014modal} to get a one-step derivation system, and we can also give a translation between $\mathbb{K}_1$-based $\mathbb{K}_2$-measure models (see the definition in \cite{cintula2014modal}) and one-step $T$-models. The concept of completeness, however, is slightly different from the one-step completeness defined in this article which makes it not an immediate instance for one-step sound and complete many-valued modal logic. Therefore, we left it as an open problem to find a one-step sound and complete many-valued modal logic.

Finally, according to \cite{kupke2015weak}, there exists a sound and weak complete logical system for coalgebraic dynamic logic. The proof was using predicate lifting and one-step approach. Besides, the soundness and weak completeness of finitely-valued propositional dynamic logic has also been proved recently\cite{Finitevalued2020}. Therefore, one can try to generalize the methods in this paper to prove the soundness and completeness of many-valued coalgebraic dynamic logic. This is another possible direction of research for future study.

\section*{Acknowledgments}
We would like to thank the Ministry of Science and Technology, Taiwan for its financial support (Grant No. MOST 110-2221-E-001-022-MY3).

\bibliographystyle{plain}
\bibliography{bibfile.bib}

\begin{thebibliography}{10}

\bibitem{Awodey}
S.~Awodey.
\newblock {\em Category Theory}.
\newblock Oxford University Press, 2nd edition, 2010.

\bibitem{baaz}
M.~Baaz.
\newblock Infinite-valued {G}\"odel logics with {$0$}-{$1$}-projections and
  relativizations.
\newblock In P.~H\'{a}jek, editor, {\em G\"odel'96: Logical Foundations of
  Mathematics, Computer Science, and Physics}, volume~6 of {\em Lecture Notes
  in Logic}, pages 23--33. Springer, 1996.

\bibitem{baldi2020classical}
P.~Baldi, P.~Cintula, and C.~Noguera.
\newblock Classical and fuzzy two-layered modal logics for uncertainty:
  Translations and proof-theory.
\newblock {\em International Journal of Computational Intelligence Systems},
  13(1):988--1001, 2020.

\bibitem{bilkova2013many}
M.~B{\'\i}lkov{\'a} and M.~Dost{\'a}l.
\newblock Many-valued relation lifting and {M}oss' coalgebraic logic.
\newblock In {\em International Conference on Algebra and Coalgebra in Computer
  Science}, pages 66--79. Springer, 2013.

\bibitem{bilkova2016expressivity}
M.~B{\'\i}lkov{\'a} and Ma. Dost{\'a}l.
\newblock Expressivity of many-valued modal logics, coalgebraically.
\newblock In {\em International Workshop on Logic, Language, Information, and
  Computation}, pages 109--124. Springer, 2016.

\bibitem{blak}
P.~Blackburn, M.~de~Rijke, and Y.~Venema.
\newblock {\em Modal Logic}.
\newblock Cambridge University Press, Cambridge, United Kingdom, 2001.

\bibitem{bou2011minimum}
F.~Bou, F.~Esteva, L.~Godo, and R.O. Rodriguez.
\newblock On the minimum many-valued modal logic over a finite residuated
  lattice.
\newblock {\em Journal of Logic and computation}, 21(5):739--790, 2011.

\bibitem{bur}
J.P. Burgess.
\newblock A quick completeness proofs for some logics of conditionals.
\newblock {\em Notre Dame J. of Formal Logic}, 22(1):76--84, 1981.

\bibitem{fml13}
X.~Caicedo, G.~Metcalfe, R.O. Rodr\'{\i}guez, and J.~Rogger.
\newblock A finite model property for {G}{\"o}del modal logics.
\newblock In L.~Libkin, U.~Kohlenbach, and R.J.G.B. de~Queiroz, editors, {\em
  Proceedings of the 20th International Workshop on Logic, Language,
  Information, and Computation (WoLLIC)}, LNCS 8071, pages 226--237.
  Springer-Verlag, 2013.

\bibitem{fml10}
X.~Caicedo and R.O. Rodr\'{\i}guez.
\newblock Standard {G}{\"o}del modal logics.
\newblock {\em Studia Logica}, 94(2):189--214, 2010.

\bibitem{fml12}
X.~Caicedo and R.O. Rodr\'{\i}guez.
\newblock Bi-modal {G}{\"o}del logic over [0,1]-valued {K}ripke frames.
\newblock {\em Journal of Logic and Computation}, 25(1):37--55, 2015.

\bibitem{che}
B.F. Chellas.
\newblock {\em Modal Logic : An Introduction}.
\newblock Cambridge University Press, 1980.

\bibitem{Cintula19}
P.~Cintula, P.~Menchón, and C.~Noguera.
\newblock Toward a general frame semantics for modal many-valued logics.
\newblock {\em Soft Computing}, 23(7):2233--2241, 2019.

\bibitem{cintula2014modal}
P.~Cintula and C.~Noguera.
\newblock Modal logics of uncertainty with two-layer syntax: A general
  completeness theorem.
\newblock In {\em International Workshop on Logic, Language, Information, and
  Computation}, pages 124--136. Springer, 2014.

\bibitem{Cintula18}
P.~Cintula and C.~Noguera.
\newblock Neighborhood semantics for modal many-valued logics.
\newblock {\em Fuzzy Sets and Systems}, 345:99--112, 2018.

\bibitem{fitting1991many}
M.~Fitting.
\newblock Many-valued modal logics.
\newblock {\em Fundamenta Informaticae}, 15(3-4):235--254, 1991.

\bibitem{fitting1992many}
M.~Fitting.
\newblock Many-valued model logics {II}.
\newblock {\em Fundamenta Informaticae}, 17(1-2):55--73, 1992.

\bibitem{hajek2013metamathematics}
P.~H{\'a}jek.
\newblock {\em Metamathematics of fuzzy logic}.
\newblock Springer, 1998.

\bibitem{Hajek07}
P.~H{\'{a}}jek.
\newblock Complexity of fuzzy probability logics {II}.
\newblock {\em Fuzzy Sets and Systems}, 158(23):2605--2611, 2007.

\bibitem{hajek10}
P.~H{\'a}jek.
\newblock On fuzzy modal logics {S5(L)}.
\newblock {\em Fuzzy Sets and Systems}, 161(18):2389--2396, 2010.

\bibitem{hajekGE95}
P.~H{\'{a}}jek, L.~Godo, and F.~Esteva.
\newblock Fuzzy logic and probability.
\newblock In P.~Besnard and S.~Hanks, editors, {\em Proceedings of the Eleventh
  Annual Conference on Uncertainty in Artificial Intelligence ({UAI}'95)},
  pages 237--244. Morgan Kaufmann, 1995.

\bibitem{halpern17}
J.Y. Halpern.
\newblock {\em Reasoning about Uncertainty}.
\newblock The MIT Press, 2 edition, 2017.

\bibitem{jacobs2017introduction}
B.~Jacobs.
\newblock {\em Introduction to Coalgebra}, volume~59.
\newblock Cambridge University Press, 2017.

\bibitem{kupke2015weak}
C.~Kupke and H.H. Hansen.
\newblock Weak completeness of coalgebraic dynamic logics.
\newblock In {\em Proceedings of the 10th International Workshop on Fixed
  Points in Computer Science (FICS 2015)}, pages 90--104, 2015.

\bibitem{kupke2011coalgebraic}
C.~Kupke and D.~Pattinson.
\newblock Coalgebraic semantics of modal logics: an overview.
\newblock {\em Theoretical Computer Science}, 412(38):5070--5094, 2011.

\bibitem{maruyama11}
Y.~Maruyama.
\newblock Reasoning about fuzzy belief and common belief: With emphasis on
  incomparable beliefs.
\newblock In {\em Proceedings of the 22nd International Joint Conference on
  Artificial Intelligence ({IJCAI})}, pages 1008--1013, 2011.

\bibitem{lmcs11}
G.~Metcalfe and N.~Olivetti.
\newblock Towards a proof theory of {G}{\"o}del modal logics.
\newblock {\em Logical Methods in Computer Science}, 7(2):1--27, 2011.

\bibitem{moss1999coalgebraic}
L.S. Moss.
\newblock Coalgebraic logic.
\newblock {\em Annals of Pure and Applied Logic}, 96(1-3):277--317, 1999.

\bibitem{ognjanovic16}
Z.~Ognjanovi\'{c}, M.~Ra\u{s}kovi\'{c}, and Z.~Markovi\'{c}.
\newblock {\em Probability Logics: Probability-Based Formalization of Uncertain
  Reasoning}.
\newblock Springer, 2016.

\bibitem{Ono2003}
H.~Ono.
\newblock Substructural logics and residuated lattices --- an introduction.
\newblock In V.F. Hendricks and J.~Malinowski, editors, {\em Trends in Logic:
  50 Years of Studia Logica}, pages 193--228. Springer Netherlands, 2003.

\bibitem{pacuit}
E.~Pacuit.
\newblock {\em Neighborhood Semantics for Modal Logic}.
\newblock Springer, 2017.

\bibitem{pattinson2003coalgebraic}
D.~Pattinson.
\newblock Coalgebraic modal logic: Soundness, completeness and decidability of
  local consequence.
\newblock {\em Theoretical Computer Science}, 309(1-3):177--193, 2003.

\bibitem{schroder2007finite}
L.~Schr{\"o}der.
\newblock A finite model construction for coalgebraic modal logic.
\newblock {\em The Journal of Logic and Algebraic Programming},
  73(1-2):97--110, 2007.

\bibitem{SchroderP09}
L.~Schr{\"{o}}der and D.~Pattinson.
\newblock Strong completeness of coalgebraic modal logics.
\newblock In S.~Albers and J.Y. Marion, editors, {\em Proceedings of the 26th
  International Symposium on Theoretical Aspects of Computer Science}, volume~3
  of {\em LIPIcs}, pages 673--684. Schloss Dagstuhl - Leibniz-Zentrum f{\"{u}}r
  Informatik, Germany, 2009.

\bibitem{schroder2010rank}
L.~Schr{\"o}der and D.~Pattinson.
\newblock Rank-1 modal logics are coalgebraic.
\newblock {\em Journal of Logic and Computation}, 20(5):1113--1147, 2010.

\bibitem{SchroderP11}
L.~Schr{\"{o}}der and D.~Pattinson.
\newblock Description logics and fuzzy probability.
\newblock In {\em Proceedings of the 22nd International Joint Conference on
  Artificial Intelligence ({IJCAI})}, pages 1075--1081. AAAI, 2011.

\bibitem{Finitevalued2020}
I.~Sedl{\'a}r.
\newblock Finitely-valued propositional dynamic logics.
\newblock In {\em Advances in Modal Logic}, pages 561--579. College
  Publications, 2020.

\bibitem{hoek92}
W.~van~der Hoek.
\newblock On the semantics of graded modalities.
\newblock {\em Journal of Applied Non-Classical Logics}, 2(1):81--123, 1992.

\bibitem{venema2012completeness}
Y.~Venema, A.~Kurz, and C.~Kupke.
\newblock Completeness for the coalgebraic cover modality.
\newblock {\em Logical Methods in Computer Science}, 8:1--76, 2012.

\end{thebibliography}

\appendix

\section{Basic Notions of Category Theory}
In this appendix, we review some basic notions of category by following the presentation in \cite{Awodey}.
\begin{defn}
A category $\mathsf{C}$ consists of the following components
\begin{itemize}
	\item A collection of {\em objects}: $Ob(\mathsf{C})=\{A, B, C, \cdots\}$
	\item A collection of {\em arrows (morphisms)}: $Ar(\mathsf{C})=\{f, g, h, \cdots\}$
	\item For each arrow $f$, there are given objects $A=dom(f)$ and $B=cod(f)$, called the domain and codomain of $f$ respectively. We use $f:A\to B$ to indicate an arrow with its {\em domain} and {\em codomain} at the same time.
	\item For arrows $f:A\to B$ and $g:B\to C$, there exists an arrow $g\circ f:A\to C$, called the {\em composite} of $f$ and $g$.
	\item For each object $A$, there is an arrow $1_A:A\to A$, called the {\em identity arrow} of $A$.
	\item Identity arrows and composites are required to satisfy the following laws:
	\begin{itemize}
		\item Associativity: for all $f:A\to B, g:B\to C$, and $h:C\to D$,\[h\circ (g\circ f)=(h\circ g)\circ f\]
		\item Unit: for all $f:A\to B$\[f\circ 1_A=f=1_B\circ f.\]
	\end{itemize}	
\end{itemize}
\end{defn}
We use $\mathsf{C}, \mathsf{D}$, etc.\ to denote a category. A particular example of category used in this paper is $\mathsf{Set}$, whose objects and morphisms are simply sets and functions respectively. The composite of two functions is their functional composition and the identity arrow of a set is the identity function on it. The {\em opposite\/} category $\mathsf{C}^{op}$ of a category $\mathsf{C}$ has the same objects as $\mathsf{C}$ and an arrow $f:C\to D$ in $\mathsf{C}^{op}$ is an arrow $f:D\to C$ in $\mathsf{C}$.

Given two objects $A$ and $B$ in any category $\mathsf{C}$, we write $Hom(A,B)=\{f\in Ar(\mathsf{C})\mid f:A\to B\}$ and call such a set of arrows as a Hom-set. Note that Hom-set is an object of the category $\mathsf{Set}$ by definition.

Just like functions play the role of arrows between sets, there is a corresponding notion of mappings between categories.
\begin{defn}
A functor $F:\mathsf{C}\to\mathsf{D}$ between two  categories $\mathsf{C}$ and $\mathsf{D}$ is defined as a mapping of objects to objects and arrows to arrows such that
\begin{itemize}
	\item if $f:A\to B$ is an arrow in $Ar(\mathsf{C})$, then $F(f):F(A)\to F(B)$ is an arrow in $Ar(\mathsf{D})$,
	\item for arrows $f:A\to B$ and $g:B\to C$ in $Ar(\mathsf{C})$, $F(g\circ f)=F(g)\circ F(f)$, and
	\item for each object $A\in Ob(\mathsf{C})$, $F(1_A)=1_{F(A)}$
\end{itemize}
\end{defn}
We sometimes omit the parentheses in the application of a functor to objects or arrows in case of no ambiguity. That is, we will usually write $FA$ or $Ff$ instead of $F(A)$ or $F(f)$. In addition, for the successive application of functors, we can write $FGA$ and $FGf$ to denote $F(G(A))$ and $F(G(f))$ respectively.

Because the opposite category $\mathsf{C}^{op}$  has the same objects as $\mathsf{C}$, a functor from $\mathsf{C}^{op}$ to $\mathsf{D}$ can also be regarded as a functor from $\mathsf{C}$ to $\mathsf{D}$ in the following sense.
\begin{defn}
A functor $F:\mathsf{C}^{op}\to\mathsf{D}$ is called a contravariant functor on $\mathsf{C}$ such that it takes each arrow $f:A\to B$ in $\mathsf{C}$ to $F(f):F(B)\to F(A)$ and satisfies that $F(g\circ f)=F(f)\circ F(g)$ for any $f:A\to B$ and $g:B\to C$ in $\mathsf{C}$.
\end{defn}
In contrast with contravariant functor, an ordinary functor is also called a {\em covariant\/} functor. A well-known example of functor is the (covariant) powerset functor ${\mathcal P}:\mathsf{Set}\to\mathsf{Set}$ and the contravariant powerset functor $\breve{\mathcal P}:\mathsf{Set}\to\mathsf{Set}$. For any set $X$, both ${\mathcal P}X$ and $\breve{\mathcal P}X$ are the powerset of $X$ and for any function $f:X\to Y$, ${\mathcal P}f:{\mathcal P}X\to {\mathcal P}Y$ and $\breve{\mathcal P}f:{\mathcal P}Y\to {\mathcal P}X$ are respectively defined by:
\[{\mathcal P}f(U)=f[U]:=\{f(x)\mid x\in U\},\;\mbox{\rm for any}\; U\subseteq X\]
and
\[\breve{\mathcal P}f(V)=f^{-1}[V]:=\{x\in X\mid f(x)\in V\},\;\mbox{\rm for any}\; V\subseteq Y.\]

For fixed categories $\mathsf{C}$ and $\mathsf{D}$, we can consider functors between them as objects of a new category. Then, the arrows between these functors (i.e.\ objects in the new category) are called natural transformations. Formally, we have the following definition.
\begin{defn}
Let $F, G:\mathsf{C}\to\mathsf{D}$ be two functors between categories $\mathsf{C}$ and $\mathsf{D}$. Then, a natural transformation $\vartheta: F\Rightarrow G$ is a family of arrows in $Ar(\mathsf{D})$, indexed by objects in $Ob(\mathsf{C})$, denoted by
\[(\vartheta_C: F(C)\to G(C))_{C\in Ob(\mathsf{C})},\]
such that, for any $f:C\to C'$ in $Ar(\mathsf{C})$, \[\vartheta_{C'}\circ F(f)= G(f)\circ \vartheta_C.\]
In diagram, this means the following commutativity:
\center{
\begin{tikzcd}
F(C)\arrow[r,"\vartheta_C"] \arrow[d,"F(f)"]& G(C) \arrow[d,"G(f)"]\\
F(C')\arrow[r,"\vartheta_{C'}"] & G(C')\\
\end{tikzcd}}
\end{defn}
Given a natural transformation $\vartheta: F\Rightarrow G$, the arrow $\vartheta_C\in Ar(\mathsf{D})$ is called the component of $\vartheta$ at $C$.
\section{Examples of Many-Valued Coalgebraic Modal Logic}
As many-valued coalgebraic modal logic provides an uniform framework for a variety of many-valued modal logics, we instantiate it to some specific examples in this appendix.
\begin{eg}
We can define the crisp Kripke model for many-valued modal logic (\cite{bou2011minimum}) coalgebraically by using the powerset functor $\mathcal P$. Let $\Lambda=\{\Box,\Diamond\}$ be the set of predicate liftings and let $\langle W,\sigma, V\rangle$ be a $\mathcal P$-model. The predicate liftings $\Box,\Diamond: Hom(W,A)\to Hom({\mathcal P}W,A)$ are defined by
\[\Box(f)(X)=\bigwedge_{x\in X}f(x)\]
\[\Diamond(f)(X)=\bigvee_{x\in X}f(x)\]
for any $f:W\to A$ and $X\subseteq W$.  Then, $\sigma:W\to{\mathcal P}W$ corresponds to the functional representation of the binary accessibility relation on $W$ and by definition, the interpretation of modal formulas is
\[\norm{\Box\varphi}_\sigma(w)=\bigwedge_{u\in \sigma(w)}\norm{\varphi}_\sigma(u)\]and
\[\norm{\Diamond\varphi}_\sigma(w)=\bigvee_{u\in \sigma(w)}\norm{\varphi}_\sigma(u),\]
which indeed correspond to the original semantics given in \cite{bou2011minimum}.
\end{eg}

\begin{eg}
Let $H$ be the Hom-functor defined in Section~\ref{sec5} and we still consider modalities in $\Lambda=\{\Box,\Diamond\}$. Then, an $H$-model $\langle W,\sigma, V\rangle$ is defined such that $\sigma:W\to Hom(W,A)$ is the functional representation of the $A$-valued accessibility relation on $W$. The predicate liftings $\Box,\Diamond: Hom(W,A)\to Hom(HW,A)$ are defined by
\[\Box(f)(g)=\bigwedge_{x\in W}g(x)\to f(x)\]
\[\Diamond(f)(g)=\bigvee_{x\in W}g(x)\odot f(x)\]
for any $f, g:W\to A$.  Hence, the interpretation of modal formulas is
\[\norm{\Box\varphi}_\sigma(w)=\bigwedge_{u\in W}(\sigma(w)(u)\to \norm{\varphi}_\sigma(u))\]and
\[\norm{\Diamond\varphi}_\sigma(w)=\bigvee_{u\in W}(\sigma(w)(u)\odot\norm{\varphi}_\sigma(u)),\]
which is exactly the same as the Kripke semantics in \cite{bou2011minimum}.
\end{eg}

\begin{eg}
We can characterize the neighborhood semantics in the coalgebraic setting by using the functor $H^2$ (i.e. the composition of the functor $H$ with itself). In an $H^2$-model $\langle W,\sigma, V\rangle$,  $\sigma:W\to Hom(Hom(W,A),A)$ is exactly the $A$-valued neighborhood function defined in \cite{Cintula18}. We only consider a modality $\Box$ because for $A$-valued neighborhood semantics, different modalities are interpreted in the same way but with different neighborhood functions. The predicate lifting $\Box: Hom(W,A)\to Hom(H^2W,A)$ is defined by
\[\Box(f)(N)=N(f)\] for any $f:W\to A$ and $N:(W\to A)\to A$. Hence, the interpretation of $\Box\varphi$ is
\[\norm{\Box\varphi}_\sigma(w)=\sigma(w)(\norm{\varphi}_\sigma),\]
precisely as that given in \cite{Cintula18}. It is easy to extend the framework to deal with multiple modalities at the same time by using the product of functors. For example, we can replace the functor $H^2$ with $H^2\times H^2$ to give semantics for both $\Box$ and $\Diamond$.
\end{eg}

\begin{eg}
The conditional logic based on the selection function semantics\cite{che} has been also presented with the coalgebraic framework in \cite{SchroderP09}. Here, we use an analogous approach to generalize it to many-valued conditional logic. Let us define the selection functor ${\mathcal S}:{\mathsf Set}\to {\mathsf Set}$ that maps a set $X$ to the set $HX\to HX$ of fuzzy selection functions and a function $f:X\to Y$ to ${\mathcal S}f:{\mathcal S}X\to{\mathcal S}Y$ defined by\footnote{We include the definition here only for the sake of completeness. It actually does not play a role in the definition of the semantics.} \[{\mathcal S}f(s)(g)(y)=\bigvee_{x:f(x)=y}s(g\circ f)(x),\]
for any $s:HX\to HX$, $g:Y\to A$, and $y\in Y$. Now, in an ${\mathcal S}$-model $\langle W,\sigma, V\rangle$, $\sigma:W\to {\mathcal S}W$ associates with each possible world $w$ a selection function $\sigma(w):Hom(W,A)\to Hom(W,A)$. Then, we consider a binary modality $\rhd$. To define the predicate lifting for the modality, we first recall the standard definition of the degree of inclusion between two $A$-valued fuzzy sets $f,g\in Hom(W,A)$ as \[f\subseteq g:=\bigwedge_{w\in W}(f(w)\to g(w))\] and then, the predicate lifting $\rhd: Hom(W,A)\times Hom(W,A)\to Hom({\mathcal S}W,A)$ is defined by
\[\rhd(f,g)(s)=(s(f)\subseteq g).\] As a result, the interpretation of the conditional formula is
\[\norm{\varphi\rhd\psi}_\sigma(w)=\sigma(w)(\norm{\varphi}_\sigma)\subseteq\norm{\psi}_\sigma.\]
\end{eg}

\begin{eg}
By using the distribution functor, it is also possible to accommodate probabilistic reasoning about fuzzy events~\cite{Hajek07,hajekGE95} in the framework of many-valued coalgebraic modal logic~\cite{SchroderP11}. To do probability calculation, we assume that the domain of truth values $A$ can be embedding into the unit interval $[0,1]$ so that we can meaningfully combine the probability and truth values by arithmetic operations. The distribution functor ${\mathcal D}:{\mathsf Set}\to {\mathsf Set}$ maps a set $X$ to ${\mathcal D}X:=\{\mu:X\to[0,1]\mid\sum_{x\in X}\mu(x)=1\}$ (i.e., the set of all discrete probability distributions over $X$) and a function $f:X\to Y$ to ${\mathcal D}f:{\mathcal D}X\to{\mathcal D}Y$ such that
\[{\mathcal D}f(\mu)(y)=\sum_{x:f(x)=y}\mu(x)\]
for any $\mu\in{\mathcal D}X$.  Then, for a ${\mathcal D}$-model $\langle W,\sigma, V\rangle$, $\sigma:W\to {\mathcal D}W$ assigns to each possible world a probability distribution over $W$. We exemplify two instances of probability modalities in the coalgebraic framework, $\mathbf P$ and ${\mathbf M}_r$ where $r\in[0,1]$ is a rational number, meaning ``probably'' and ``withe probability more than r'' respectively~\cite{SchroderP11}. The predicate liftings of these modalities ${\mathbf P}, {\mathbf M}_r: Hom(W,A)\to Hom({\mathcal D}W,A)$ are defined as follows:
\[{\mathbf P}(f)(\mu)=\sum_{x\in W}f(x)\cdot\mu(x),\]
\[{\mathbf M}_r(f)(\mu)=\bigvee\{\alpha\mid\mu(f_\alpha)>r\},\]
for any $f:W\to A$ and $\mu\in{\mathcal D}W$, where $f_\alpha:=\{x\in W\mid f(x)\geq\alpha\}$ is the $\alpha$-cut of $f$ (regarded as an $A$-valued fuzzy set).  Thus, by definition, the interpretation of probability modal formulas is
\[\norm{{\mathbf P}\varphi}_\sigma(w)=\sum_{x\in W}\norm{\varphi}_\sigma(x)\cdot\sigma(w)(x),\]
\[\norm{{\mathbf M}_r\varphi}_\sigma(w)=\bigvee\{\alpha\mid\sigma(w)((\norm{\varphi}_\sigma)_\alpha)>r\},\] where $\norm{\varphi}_\sigma: W\to A$ is regarded as an $A$-valued fuzzy set.
\end{eg}

\end{document}